# Double band inversion in the topological phase transition of Ge$_{1-x}$Sn$_x$ alloys

Rui Tan[1], Qirui Yang[1], Ruiyuan Chen[1], Yu Qiao[1], Shouqiao Tan[1], Peng Wang[2], Tai-Chang Chiang[3,4], Xiaoxiong Wang[1]

[1] *Department of Applied Physics, Nanjing University of Science and Technology, Nanjing 210094, China*
[2] *College of Electronic, Communication and Physics, Shandong University of Science and Technology, Qingdao 266590, China*
[3] *Department of Physics, University of Illinois at Urbana-Champaign, 1110 West Green Street, Urbana, Illinois 61801-3080, USA*
[4] *Frederick Seitz Materials Research Laboratory, University of Illinois at Urbana-Champaign, 104 South Goodwin Avenue, Urbana, Illinois 61801-2902, USA*

**Abstract** – We use first-principles simulation and virtual crystal approximation to reveal the unique double band inversion and topological phase transition in Ge$_{1-x}$Sn$_x$ alloys. Wavefunction parity, spatial charge distribution and surface state spectrum analyses suggest that the band inversion in Ge$_{1-x}$Sn$_x$ is relayed by its first valence band. As the system evolves from Ge to α-Sn, its conduction band moves down, and inverts with the first and the second valence bands consecutively. The first band inversion makes the system nontrivial, while the second one does not change the topological invariant of the system. Both the band inversions yield surface modes spanning the individual inverted gaps, but only the surface mode in the upper gap associates with the nontrivial nature of tensile-strained α-Sn.

**Introduction.** – Over the past decades topological phases and topological materials have continually attracted tremendous attention from the condensed matter physics community for interesting fundamental physics and great potential applications therein [1-4]. Based on the topology of their band structures [5], materials can be classified into a few categories, such as topological insulator [1], topological Dirac semimetal [6], Weyl semimetal [7], etc. It is intriguing to explore and engineer the transformation of materials between distinct topological phases. Researchers have devoted many efforts to investigate the topological phase transition on various topological materials, such as Bi$_{1-x}$Sb$_x$ [8], Sb$_2$Se$_3$ [9], GaS and GaSe [10], TlBi(S$_{1-x}$Se$_x$)$_2$ [11], (Bi$_{1-x}$In$_x$)$_2$Se$_3$ [12] and Pb$_{1-x}$Sn$_x$Te [13]. The underlaying physics behind all these diverse systems is to tune lattice constant by strain or modulate effective spin-orbit coupling strength by incorporating foreign atoms, usually in the same column of the periodic table, into the host lattice to realize band inversion and topological phase transition. As the strain or composition varies, the fundamental gap slowly reduces to zero. With further tuning the parameters, the gap reopens with inverted band order and the system transforms into or out of a topological phase. Usually, the band inversion occurs between the lowest conduction band and the highest valence band with opposite parity, for example in the case of the stereotype topological insulator, Bi$_2$Se$_3$ [14].

α-Sn is different from the common topological materials. It crystallizes in a diamond-like crystal structure with cubic symmetry. Compared with germanium, α-Sn is topologically nontrivial because of its inverted *s, p* bands [15], namely, its conduction band and second valence band are inverted. In equilibrium geometry, its first valence band traverses the inverted *sp* gap and touches the conduction band by a single point at the zone center as constraint by the cubic symmetry; α-Sn is a topological semimetal. By applying an appropriate in-plane tensile strain to break the cubic symmetry and lift the protected degeneracy between the conduction and the first valence bands, a small global gap around the Fermi level can be obtained; α-Sn transforms into a topological insulator [15,16]. By applying a compressive strain, α-Sn can be tuned into a Dirac semimetal or even a Weyl semimetal if its time-reversal symmetry is broken with a magnetic field [17]. α-Sn is an ideal platform for studying topological phase transition and engineering for its rich topological properties.

Although many groups have theoretically and experimentally studied the electronic properties of α-Sn [15-18], the question about the role played by the first valence band in the band inversion and topological phase transition is seldomly addressed and deserves an in-depth study. The photoemission spectroscopy results also needed to be revisited to reach a consistent understanding on the theoretical and experimental results. According to the prior studies, the band inversion of α-Sn occurs between the conduction *s* band and the second valence *p* band. It seems the first valence *p* band is just a spectator rather than a player, and has nothing to do with the band inversion and topological phase transition [18]. The topological surface state connects the inverted conduction band and the second valence band; the impact of the first valence band on the topological surface states is supposed to be small as implied in the model proposed by Barfuss *et al*

[18]. The interaction between the topological surface states and the first valence band states may result in hybridization between them and make the latter surface localized and spin polarized to some extent. Otherwise, the intact topological surface states will cross the inverted *sp* gap. However, the calculations on α-Sn films show that the states of the first valence band do not show any sign of surface localization and spin polarization. Reported angle-resolved photoemission spectroscopy (ARPES) measurements show that the Dirac cone of α-Sn is deeply buried in its valence band continuum. Based on our knowledge, if the Dirac cone moves into the bulk band continuum, it will be absorbed by the bulk bands.

To clarify all the above controversies, we use first-principles simulation to illustrate the band inversion and topological phase transition in $Ge_{1-x}Sn_x$ alloys and to understand the unique band structure and topological properties of α-Sn. The results clearly show a double band inversion, the conduction band invert with the first and the second valence bands consecutively. The first inversion pushes the system into a nontrivial state, while the second one does not change the topological nature of the system. The surface state spectrum calculations on the alloys with critical compositions also corroborate our statements.

**Computational methods.** – The first-principle calculations were based on density-functional theory (DFT) as implemented in the Vienna ab initio simulation package (VASP) package [19] and Abinit package [20]. In the VASP calculations, projector augmented-wave (PAW) pseudopotential [21] was used; the Abinit calculations were done with the Hartwigsen-Goedecker-Hutter (HGH) relativistic separable dual-space Gaussian pseudopotential [22]. The lattice parameters of Ge and Sn were fully relaxed by VASP using local density approximation (LDA), while the lattice parameters of $Ge_{1-x}Sn_x$ alloy were linearly interpolated based on the value x and the lattice parameters of Ge and Sn. The criterion for geometry relaxation was that the Hellmann–Feynman force on each atom is less than 0.005 eV/Å. Some $Ge_{1-x}Sn_x$ substitutional alloys were relaxed based on supercell models to verify the applicability of the linear interpolation method in predicting the lattice parameters of $Ge_{1-x}Sn_x$ alloys. The resultant lattice parameters are consistent with the linear prediction well. The band structures of pure Ge, Sn crystal were calculated with VASP if not otherwise declared, and those of $Ge_{1-x}Sn_x$ alloys were calculated with Abinit using virtual crystal approximation [23,24]. The bulk energy spectra were corrected with modified Becke-Johnson potential (mBJLDA) correction [25]. The kinetic energy cutoff was 500 eV for plane wave basis. The integration in momentum space was performed on a Γ-centred 11×11×5 grid [26]. The surface state of $Ge_{1-x}Sn_x$ alloys was calculated using a tight-binding model and green function method [27,28]. The hoping parameters were determined by fitting the bulk energy band with a $sp^3$ basis.

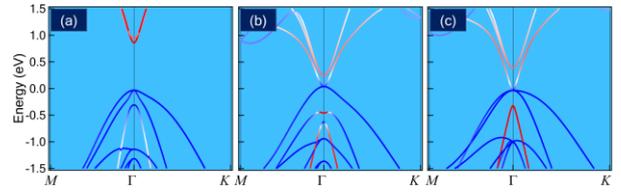

Fig. 1: (Color online) Band structures of Ge and Sn in equilibrium geometry. (a) Ge, (b) α-Sn, (c) α-Sn without SOC. The bands are colored based on the weight of *s* state. Red color represents *s* state, blue represents *p* state.

**Results and discussion.** – Fig. 1 presents the band structures of pristine bulk Ge and Sn. To show the orbital character of the states of interest, an orbital decomposition was performed. The red color represents the weight of *s*-state, blue represents *p* state. Apparently, the conduction band of Ge [fig.1 (a)] is *s*-state like, while the two top valence bands are *p*-state like. Due to cubic symmetry, the two *p* valence bands are degenerate at the zone center. As for the band structure of α-Sn [fig.1 (b)], its conduction band bottom is obviously of *p*-state characteristic. The conduction band and valence band touch with each other by a single point. This degenerate point is protected by cubic crystal symmetry. The states of the second valence band at the zone center are *s*-state like, and there seems a band crossing between the second and the third valence bands at Γ point. Compared with the band structure of Ge, that of α-Sn has gone through a band inversion between the *s* (the conduction band) and *p* (the second valence band) states. Because these two states have opposite parity, this inversion makes the band structure of α-Sn topologically nontrivial. α-Sn of cubic symmetry is a topological semimetal. The band structure of α-Sn without spin-orbit coupling (SOC) is shown in fig. 1(c). Interestingly, even when SOC is turned off, the *s*, *p* bands are inverted as the case with SOC [fig. 1(b)]. So, based on the parity argument, α-Sn is topologically nontrivial no matter whether SOC is counted in or not. This is different from the general cases such as $Bi_2Se_3$. In $Bi_2Se_3$, when SOC is turned off, it is topologically trivial and its band order is normal [14]. Obviously, SOC plays a crucial role in the topological nature of $Bi_2Se_3$-like topological materials. By contrast, SOC has little effect on the topological properties of α-Sn.

Sine SOC strength is not the dominant factor in determining the topological nature of α-Sn, next we check whether its lattice constant can effectively tune its topological property or not. We theoretically expand the lattice of Ge gradually to check whether it is possible to realize a topological phase transition and reproduce the nontrivial band topology of α-Sn. The results are presented in fig. 2. As the lattice of Ge expands from its equilibrium geometry to 3% [fig. 2(a)-(d)], the conduction band slowly moves downwards and the fundamental gap keeps reducing. At 3% the conduction band and the valence band basically degenerate, and the gap approaches zero. As the lattice expansion up to 4%, fig. 2(e), the degeneracy between the first and the second valence bands is lifted. This is induced by the band inversion between the conduction band edge and the maximum of the second valence band. We can see that fig. 2(e) is very similar to the band structure of α-Sn [fig. 1(b)]. Evidently, the band structure of 4% expanded Ge successfully captures the major feature of the nontrivial band structure of α-Sn. Therefore, to expand the lattice of Ge, we are expected to observe the band inversion and topological phase transition, and then understand the band inversion and topological property of α-Sn ultimately.

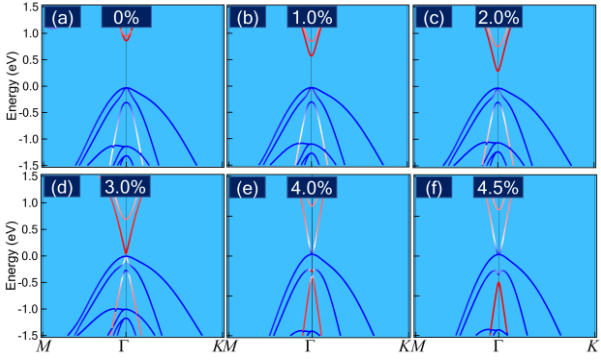

Fig. 2: (Color online) Band structures of Ge with expanded lattice as indicated. The bands are colored based on the weight of *s* orbital of the states to illustrate the inversion between *s* and *p* bands.

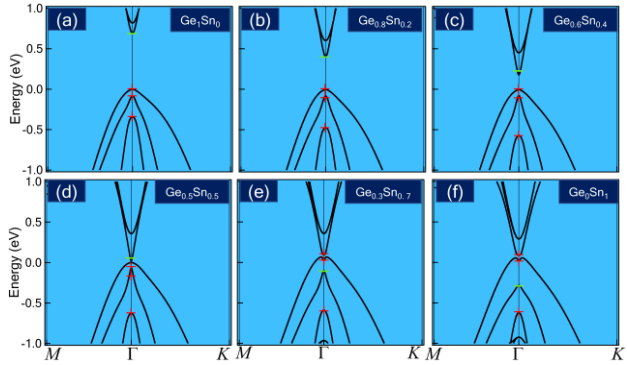

Fig. 3: (Color online) Band structure of 0.5% tensile-strained $Ge_{1-x}Sn_x$ alloys. The parity of the states at zone center is labelled.

However, it is inaccessible experimentally to expand the lattice constants of a material in three dimensions with mechanical strain simultaneously. Luckily, it is a common strategy to expand a crystal's lattice with chemical strain, namely, incorporating foreign atoms with large radius, usually in the same column of the periodic table, into the host lattice to expand its lattice. Hence, we calculated the band structures of 0.5% in-plane tensile-strained $Ge_{1-x}Sn_x$ alloys, aiming at witnessing the band inversion and topological phase transition in the alloys. A 0.5% in-plane tensile-strain can create a decent global band gap in α-Sn, which makes it possible to show the topological surface state in the gap. Meanwhile, the tensile-strain can lift the band degeneracy protected by cubic symmetry and ambiguously show the band inversion details. The lattice constant of the alloys was interpolated linearly based on the lattice parameters of 0.5% tensile-strained Ge and Sn. The *c* axes of the 0.5% tensile-strained Ge and Sn were fully relaxed with VASP. Geometry relaxation, assuming there is no structure phase transition, on some representative substitutional $Ge_{1-x}Sn_x$ alloys suggests that linear approximation is good enough to predict the lattice constants of $Ge_{1-x}Sn_x$ alloys [see supplementary materials]. We used a virtual crystal approximation [23,24] to calculate the band structures of $Ge_{1-x}Sn_x$ alloys with various stoichiometries. The band structures of 0.5% tensile-strained $Ge_{1-x}Sn_x$ alloys from Ge through Sn calculated with Abinit are depicted in fig. 3. The band structures of unstrained $Ge_{1-x}Sn_x$ alloys are presented in the supplementary materials as a complementary. The band structures of unstrained $Ge_{1-x}Sn_x$ alloys with increasing Sn contents successfully reproduce the band structure features of Ge with expanded lattices. The band structures of strained $Ge_{1-x}Sn_x$ alloys are some kind like those of unstrained ones, except that the strain breaks the cubic symmetry and lifts the degeneracy between the conduction band and the valence band, generating a finite global band gap around the Fermi level. The finite gap makes the analysis on band inversion unambiguously. To show the band inversion clearly, the parity of the states close to the Fermi level is labelled. Fig. 3 shows a band inversion between the conduction band and the second valence band as the composition varies from $Ge_{0.5}Sn_{0.5}$ [fig. 3(d)] to $Ge_{0.3}Sn_{0.7}$ [fig. 3(e)].

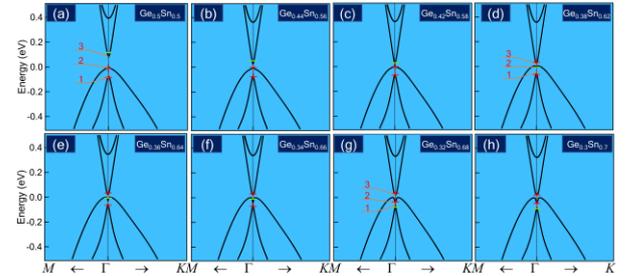

Fig. 4: (Color online) The detailed band structure of 0.5% tensile-strained $Ge_{1-x}Sn_x$ alloys near the topological phase transition. The parity of the states at zone center is labelled. Some states are indicated for charge density plot.

However, the detailed band evolution in between these two critical compositions was missed. So, we did a more thorough calculation on the alloys between $Ge_{0.5}Sn_{0.5}$ and $Ge_{0.3}Sn_{0.7}$ with a finer tuning on stoichiometries, the results are presented in fig. 4. As the content of Sn increases from $Ge_{0.5}Sn_{0.5}$, the position of the conduction band gradually and monotonically moves down, until it touches the first valence band at $Ge_{0.42}Sn_{0.58}$. With Sn increasing a little more [fig. 4(d)], the conduction band inverts with the first valence band at the zone center. As the parity indicated, the conduction band and the first valence band have opposite parity, so the inversion between this two bands makes the system nontrivial. Now the system is a topological insulator. When the ratio of Ge:Sn reaches 0.32:0.68 [fig. 4(g)], the first valence band and the second valence band switch band order, and the resultant band order becomes α-Sn like. The second inversion occurs between two occupied bands, so this inversion does not change the topological invariant of the system although these two bands have opposite parity; it is a trivial band inversion. In summary, as the alloys evolve from pure Ge to the Sn rich end, the conduction band moves down and switches with the first and the second valence bands, respectively. This is more clearly revealed in the supplementary materials. It is a double band inversion. The first band inversion makes the system transform into a topologically nontrivial state, while the second inversion is nothing to do with topological phase transition.

$Ge_{0.38}Sn_{0.62}$, evidently, the states 2 and 3 switch with each other, but the state 1 is intact. For $Ge_{0.32}Sn_{0.68}$, the states 1 and 2 exchange with each other. Finally, after going through a double-band-inversion, the *s* state moves to its destination, the second valence band. The results clearly illustrate again that the band inversion between *s* and *p* band is not a direct exchange, but relayed by the state 2, the first valence band. The spatial charge distribution analysis on the unstrained $Ge_{1-x}Sn_x$ alloys leads to the same conclusion as concluded above (see supplementary materials).

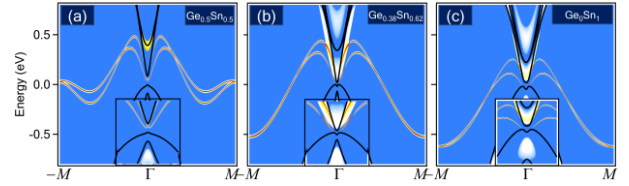

Fig. 6: (Color online) Surface state spectra overlayered with bulk band structure of some critical compositions around the topological phase transition point. The insets are the details of the band structure around the Fermi level.

The topological phase transition at the first band inversion (inversion between the conduction band and the first valence band) is also verified by the surface state spectrum calculation as presented in fig. 6. We calculated the surface state spectrum of three representative alloys, $Ge_{0.5}Sn_{0.5}$, $Ge_{0.38}Sn_{0.62}$ and $Ge_0Sn_1$, corresponding to before band inversion, after the first band inversion and after the second band inversion. For $Ge_{0.5}Sn_{0.5}$, it is prior to topological phase transition and trivial. Its surface state bands, originating from the unsaturated surface dangling bonds, reside in the fundamental band gap, and do not connect the conduction band and valence band simultaneously. The surface state of $Ge_{0.38}Sn_{0.62}$ is topologically different from that of $Ge_{0.5}Sn_{0.5}$. Its surface state bands originate from the conduction band and terminate at the first valence band edge, this is a critical feature of a topological surface state. For pure Sn, which goes through a double band inversion, its surface state spectrum is quite like that of $Ge_{0.38}Sn_{0.62}$, indicating the topological similarity between pure Sn and $Ge_{0.38}Sn_{0.62}$. We notice that there is a dim sign of surface state in the inverted gap of the first and the second valence bands. Our first-principle calculation based on a thick slab model with bare surfaces suggests that there is a Dirac cone with its upper branch connecting to the first valence band and lower branch to the second valence band (see the supplementary materials). This Dirac cone should be due to the inversion between these two bands, but it is not an indication of the nontrivial nature of α-Sn since the associated band inversion is not linked to a topological phase transition.

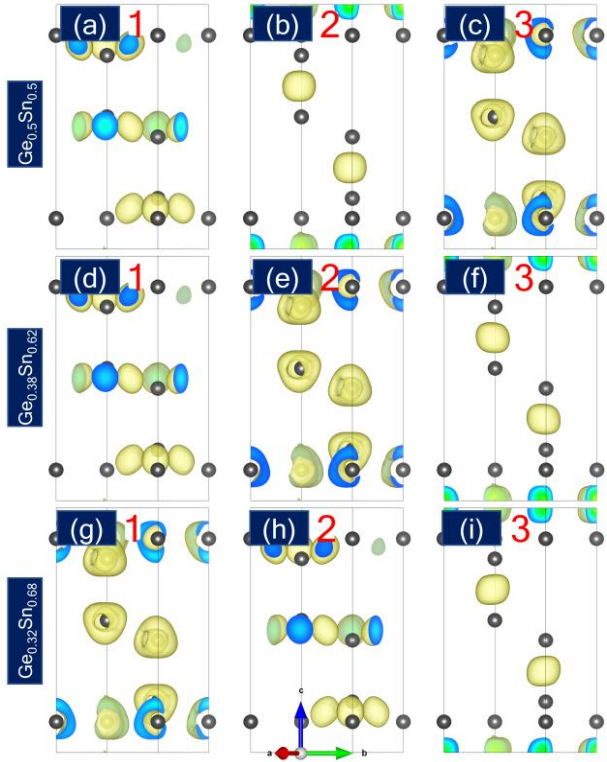

Fig. 5: (Color online) Spatial charge distribution of the states indicated in fig. 4.

The spatial charge distribution presented in fig. 5 directly supports our statement. For the composition prior to band inversion, $Ge_{0.5}Sn_{0.5}$, the conduction band bottom, state 3, is *s*-state like as shown in fig. 5(c). The charge density shows an *s*-orbital like spherical charge density distribution. The two valence bands are *p*-state like, the state 1 is $p_x+ip_y$ like, the electrons mainly localize in the plane; the state 2 is $p_z$ like, the electrons mainly distribute along the *c* direction. While for

We notice that our surface state spectrum dispersion is not consistent with previous ARPES measurement very well. This inconsistency may come from the fact that in usual ARPES measurements only the occupied band can be imaged, so the surface states in the fundamental gap is inaccessible because they are above the Fermi level and unoccupied. The second possibility is that the surface adsorbent may shift the Fermi level and distort the surface state bands dispersion, making the surface states in the global band gap even harder to be detected. The reported ARPES results show there is a Dirac cone located at about 0.5 eV below the Fermi level, we argue that this feature should be attributed to the calculated surface states in the second inverted gap, which cannot be assigned as a signature of the nontrivial topological property of α-Sn. The above statement is also supported by the fact that the observed surface states around the zone center have a much larger energy span than the size of the calculated band gap (the upper gap) even after considering HSE06 correction. This disagreement reminds us that it is necessary to revisit the previous ARPES measurements on the topological insulating α-Sn. Electron doping of α-Sn samples to shift the Fermi level upwards may let us have chance to observe the topological surface state in the fundamental gap. A scanning tunneling spectroscopy (STS) is more suitable for the present purpose, since this technique can detect both the occupied and unoccupied states.

**Concluding remarks.** – The electronic structure of $Ge_{1-x}Sn_x$ alloy was explored with first-principles simulation and virtual crystal approximation aiming at understating the topological property of α-Sn. The results illustrate that as the system evolve from Ge to Sn, its conduction band moves downwards and invert with the first and the second valence bands successively. It is a double band inversion. The first valence band is deeply involved in the band inversion, rather than a spectator. The first band inversion provokes a topological phase transition, while the second inversion leave the topological property of the system unchanged. The nontrivial band inversion associates with a nontrivial surface mode, while the trivial inversion links with a trivial surface mode.

Our findings are useful for deepening the understanding on the topological property of α-Sn and informative for the engineering and optimization of the property of $Ge_{1-x}Sn_x$ alloys as a candidate for topological material.

\*\*\*

This work received support from the National Natural Science Foundation of China (No. 11204133 for XXW), the Jiangsu Province Natural Science Foundation of China (No. BK2012393 for XXW), the U.S. Department of Energy, Office of Science (Grant DE-FG02-07ER46383 for TCC).